\documentclass[prb,twocolumn,aps,longbibliography]{revtex4-2}
\usepackage{graphicx,amsmath,xcolor}
\usepackage{hyperref}
\usepackage{subcaption}
\usepackage{float}
\usepackage{svg}
\usepackage{enumerate}

\usepackage{silence}
\usepackage{hyperref}
\usepackage{mathtools}
\usepackage[T1]{fontenc}
\DeclarePairedDelimiterX\innerp[2]{\langle}{\rangle}{#1\delimsize\vert\mathopen{}#2}%
\DeclarePairedDelimiterX\braket[2]{\langle}{\rangle}{#1\delimsize\vert\mathopen{}#2}%
\DeclarePairedDelimiterX\braketOP[3]{\langle}{\rangle}{#1\,\delimsize\vert\,\mathopen{}#2\,\delimsize\vert\,\mathopen{}#3}%
\DeclarePairedDelimiterX\ketbra[2]{\lvert}{\rvert}{#1\delimsize\rangle\!\delimsize\langle#2}%
\DeclarePairedDelimiterX\outerp[2]{\lvert}{\rvert}{#1\delimsize\rangle\!\delimsize\langle#2}%
\DeclarePairedDelimiterX\projector[1]{\lvert}{\rvert}{#1\delimsize\rangle\!\delimsize\langle#1}%
\newcommand{\comment}[1]{}
\newcommand{\subsecref}[1]{S-\Roman{section}.\arabic{subsection}}

\begin{document}
    \title{Predicting spin-orbit coupling in hole spin qubit arrays with vision-transformer-based neural networks on a generalized Hubbard model}
    \author{Jacob R. Taylor}
    \author{Katharina Laubscher}
    \author{Sankar Das Sarma}
    \affiliation{Condensed Matter Theory Center and Joint Quantum Institute, Department of Physics, University of Maryland, College Park, Maryland 20742-4111 USA}
\begin{abstract}
We introduce a neural-network-based machine learning method to predict the effective spin-orbit coupling (SOC) strength in hole quantum dot arrays from standard charge stability diagrams. Specifically, we study a $2\times 2$ Ge hole quantum dot array described by a generalized spin-orbit coupled Hubbard model that incorporates random site- and bond-dependent disorder in all system parameters, including onsite potentials, Coulomb interaction strengths, interdot tunneling amplitudes, as well as the direction and angle of the SOC-induced spin rotations accompanying interdot tunneling. We train the neural network on numerically simulated charge stability diagrams from nearest-neighbor pairs of quantum dots for different chemical potentials and out-of-plane magnetic fields, and show that this enables us to predict the SOC-induced spin-flip tunneling amplitudes---and, thus, the effective SOC strength---with high fidelity ($R^2\approx 0.94$) even when all other Hubbard model parameters are unknown. Furthermore, our neural network can also predict the other Hubbard model parameters with high fidelity, demonstrating that neural-network-based approaches can be a powerful tool for the automated characterization of hole spin qubit arrays.
\end{abstract}

\maketitle 

\textit{Introduction.\textemdash}%
Germanium (Ge) hole systems have emerged as an attractive platform for spin qubits, combining high material quality, compatibility with advanced semiconductor fabrication, and strong spin-orbit coupling (SOC) enabling fast and all-electrical qubit control~\cite{Scappucci2020}. These favorable properties have led to considerable experimental efforts, including demonstrations of coherent control of single- and two-qubit devices as well as the fabrication of larger and more complex quantum dot arrays. For example, 2$\times$2~\cite{Lawrie2020,Hendrickx2021,Lawrie2023}, 2$\times$4~\cite{Hsiao2024,Zhang2024,Farina2025,Jirovec2025}, 4$\times$4~\cite{Borsoi2022}, as well as ten-dot~\cite{John2025} and 18-dot arrays~\cite{dijkema2026} have recently been realized in gate-defined Ge/SiGe heterostructures. Beyond their direct relevance for scalable quantum computing architectures, these arrays provide a versatile platform for exploring strongly correlated solid-state phenomena under controlled conditions since quantum dot spin qubits can emulate an effective solid-state Fermi-Hubbard model~\cite{Stafford1994, Kotlyar1998,Kotlyar1998b,hensgens2017quantum,Chou2026}.

Indeed, at low energies, hole quantum dot arrays are naturally described in terms of a spin-orbit coupled extended Hubbard model, which by itself is of intrinsic interest as a highly nontrivial strongly correlated model. Here, the strong intrinsic SOC causes holes to undergo a SU(2) spin rotation upon tunneling between dots, leading to spin-flip tunneling processes in addition to the conventional spin-conserving hopping. This gives rise to qualitatively new phenomena that are absent in systems with negligible SOC, such as complex spin-spin interactions including Dzyaloshinskii–Moriya and anisotropic exchange terms. On the spin qubit side, this enables fast and robust two-qubit gates for Loss-DiVincenzo type qubits as well as novel opportunities for multi-spin encodings~\cite{Katsaros2020,Hendrickx2020,Zhang2024,SaezMollejo2025,Bosco2026,Jirovec2022,Liu2022,Kelly2025,Seidler2025}, while also crucially influencing spin shuttling protocols in sparse quantum dot arrays~\cite{vanRiggelen2024,Bosco2024}.

However, despite its qualitative importance, the effective strength of the SOC is often unknown in experiments and depends sensitively on details such as the exact device geometry and the local electrostatic environment. Although information about the SOC can be obtained by analyzing specific spectroscopic features (e.g., singlet-triplet anticrossings)~\cite{Jirovec2022,Liu2022,Kelly2025,Seidler2025}, the interpretation of these signatures becomes increasingly challenging as device complexity increases. Moreover, sample-dependent random disorder and fabrication-induced inhomogeneities render even nominally identical dots effectively distinct, such that each dot has to be considered on a case-by-case basis. This raises the question whether more systematic approaches can be developed to extract the effective SOC strength in multi-dot devices from experimental data in an automated way.

Previous work has shown that neural networks, and in particular vision transformers~\cite{dosovitskiy2020image}, provide a robust framework for solving unknown parameter problems of this type by recasting experimental measurement tasks as image-processing problems~\cite{taylor2023machine,taylor2025vision,durrer2020automated,schuff2024fully,taylor2025neural,pawlowski2026learning}. Such approaches are especially powerful when the underlying parameter space is high-dimensional and conventional fitting procedures become unstable or computationally prohibitive. The ability to identify correlations in structured, two-dimensional datasets makes vision transformers particularly promising for extracting model parameters from experimental measurements.

In this work, we demonstrate that a vision-transformer-based neural network can predict spin-flip hopping amplitudes—and thus the effective SOC strength—from standard charge stability diagrams, which are among the most routinely acquired experimental datasets in quantum dot arrays. Specifically, we describe a $2\times 2$ hole quantum dot array as a generalized spin-orbit coupled Hubbard model that incorporates random site- and bond-dependent disorder in all system parameters including onsite potentials, Coulomb interaction strengths, interdot tunneling amplitudes, as well as the direction and magnitude of the SOC. We train the neural network on simulated charge stability diagrams from nearest-neighbor pairs of quantum dots for different chemical potentials and out-of-plane magnetic fields, and show that this approach enables high-fidelity prediction of the effective SOC strength ($R^2\approx 0.94$) even when all other Hubbard model parameters are treated as unknown.

In addition to the SOC strength, our neural network can simultaneously also predict the other (conventional) Hubbard model parameters with high fidelity, consistent with recent results for systems without SOC~\cite{taylor2025neural}. This demonstrates that unknown SOC terms do not obstruct the automatic characterization of multi-dot arrays using vision transformers. Taken together, our results demonstrate that neural-network-based deep-learning approaches provide a powerful tool for the automated characterization, calibration, and fine-tuning of hole spin qubit arrays.

\begin{figure}[t]
    \centering

    \begin{subfigure}[t]{0.9\linewidth}
        \centering
        \includegraphics[width=\textwidth]{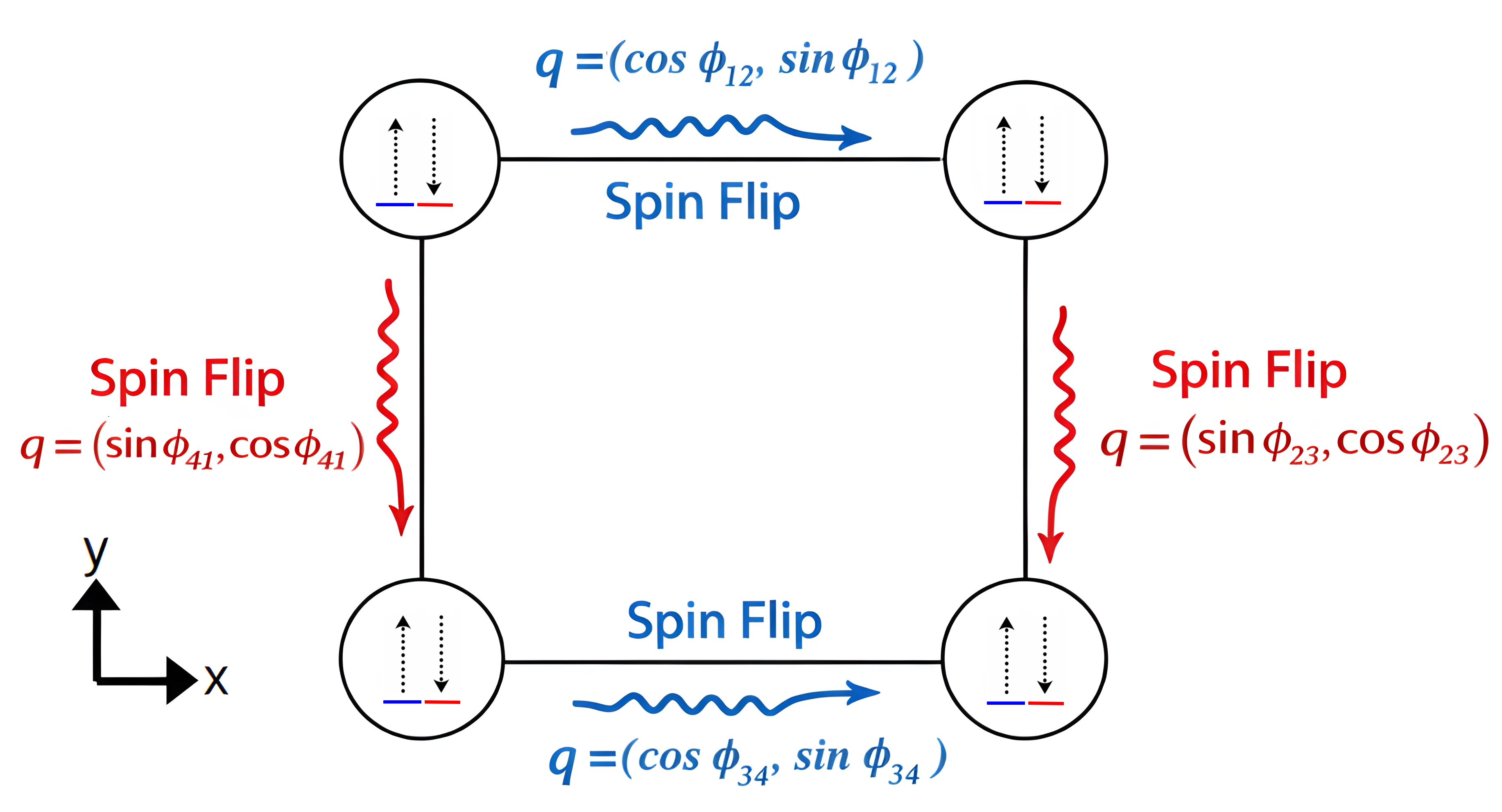}
    \end{subfigure}

    \caption{Diagram of a $2\times 2$ quantum dot array with SOC. Due to the SOC, particles undergo a SU(2) spin rotation as they tunnel between nearest-neighbor dots $i$ and $j$, leading to spin-flip hopping terms in addition to the conventional spin-conserving hopping. The corresponding axis of rotation is denoted by an in-plane vector $\vec{q_{ij}}$, parametrized by an angle $\phi_{ij}$.}
    \label{fig:LatticeDiagram}
\end{figure}

\textit{Model.\textemdash}%
We consider a 2$\times$2 quantum dot array defined in a planar Ge hole gas, see Fig.~\ref{fig:LatticeDiagram} for a schematic illustration. Such an array is well described by a generalized Hubbard model that includes spin-flip tunneling terms between neighboring dots arising from strong SOC:
\begin{align}
H&=-\sum_{\langle i,j\rangle,\alpha,\alpha'} \big[t_{ij}c^\dagger_{i\alpha}(\cos\theta_{ij}\delta_{\alpha\alpha'}+i\sin\theta_{ij}
(\vec{q}_{ij}\cdot\vec{\sigma})^{\alpha\alpha'})c_{j\alpha'}\nonumber\\&\qquad\quad+\mathrm{h.c.}\big]-\sum_i \epsilon_i n_i+\sum_i \frac{U_i}{2}\, n_i(n_i-1)\nonumber\\&\qquad\quad+\sum_{\langle i,j\rangle} V_{ij} n_i n_j + \sum_{i,\alpha,\alpha'} V_z\,c^\dagger_{i\alpha}(\sigma_z)^{\alpha\alpha'}c_{i\alpha'}.\label{eq:hubbard}
\end{align}
Here, $c_{i\sigma}^\dagger$ is the creation operator of a hole with (pseudo)spin $\sigma\in\{\uparrow,\downarrow\}$ on dot $i$, $n_i=n_{i\uparrow} + n_{i\downarrow}$ with $n_{i\sigma}=c_{i\sigma}^\dagger c_{i\sigma}$ is the number operator, $\vec{\sigma} = (\sigma_x, \sigma_y,\sigma_z)$ is the vector of Pauli spin matrices, and we use the convention that $\langle i,j\rangle$ runs over all nearest-neighbor bonds oriented along the positive lattice directions. Due to SOC, the holes undergo a SU(2) spin rotation $\exp(i\theta_{ij}\vec{q}_{ij}\cdot\vec{\sigma})=\cos\theta_{ij}+i\sin\theta_{ij}\,\vec{q}_{ij}\vec{\cdot\sigma}$ as they tunnel from dot $i$ to dot $j$, leading to spin-flip hopping terms in addition to the conventional spin-conserving hopping. Here, $\theta_{ij}$ is the angle by which the spin rotates and $\vec{q}_{ij}$ sets the axis of rotation, which we parametrize as $ \vec{q}_{ij}= (\cos\phi_{ij}, \sin\phi_{ij},0)$ [$\vec{q}_{ij} = (\sin\phi_{ij}, \cos\phi_{ij},0)$] for vertical [horizontal] bonds. With this definition, the case $\phi_{ij} = 0$ for all $i$, $j$ corresponds to standard Rashba SOC, which we will use as a simple test case for our machine learning algorithm before generalizing to arbitrary bond-dependent rotation axes. The overall hopping amplitude (experimentally controllable, e.g., by barrier gates) is denoted by $t_{ij}$. Finally, $\epsilon_i$ is the single-particle energy, $U_i$ ($V_{ij}$) is the onsite (inter-dot) Coulomb repulsion, and $V_z$ is the Zeeman splitting induced by an out-of-plane magnetic field.

Due to random unintentional disorder, which is unavoidably present in all semiconductor devices, the Hubbard model parameters are sample-dependent, and, moreover, vary spatially from dot to dot or bond to bond even in systems of nominally identical dots, making the effective model an SOC-generalized disordered Fermi-Hubbard-Mott-Anderson model~\cite{hensgens2017quantum}. We incorporate this uncertainty by randomly drawing the Hubbard model parameters from a range of values (specified below) for each site and bond separately. For simplicity, we assume that the Zeeman splitting $V_z$ is spatially uniform; however, in realistic Ge hole systems, $g$-tensor variations typically render the Zeeman splitting spatially non-uniform as well~\cite{Martinez2022,Geyer2024,Hendrickx2024,Kelly2025,Seidler2025,Valvo2025,Martinez2026}. For completeness, a generalized model including $g$-tensor variations is studied in the Supplementary Material (SM).

\begin{figure}[t]
    \centering

        \centering
        \includegraphics[width=0.9\linewidth]{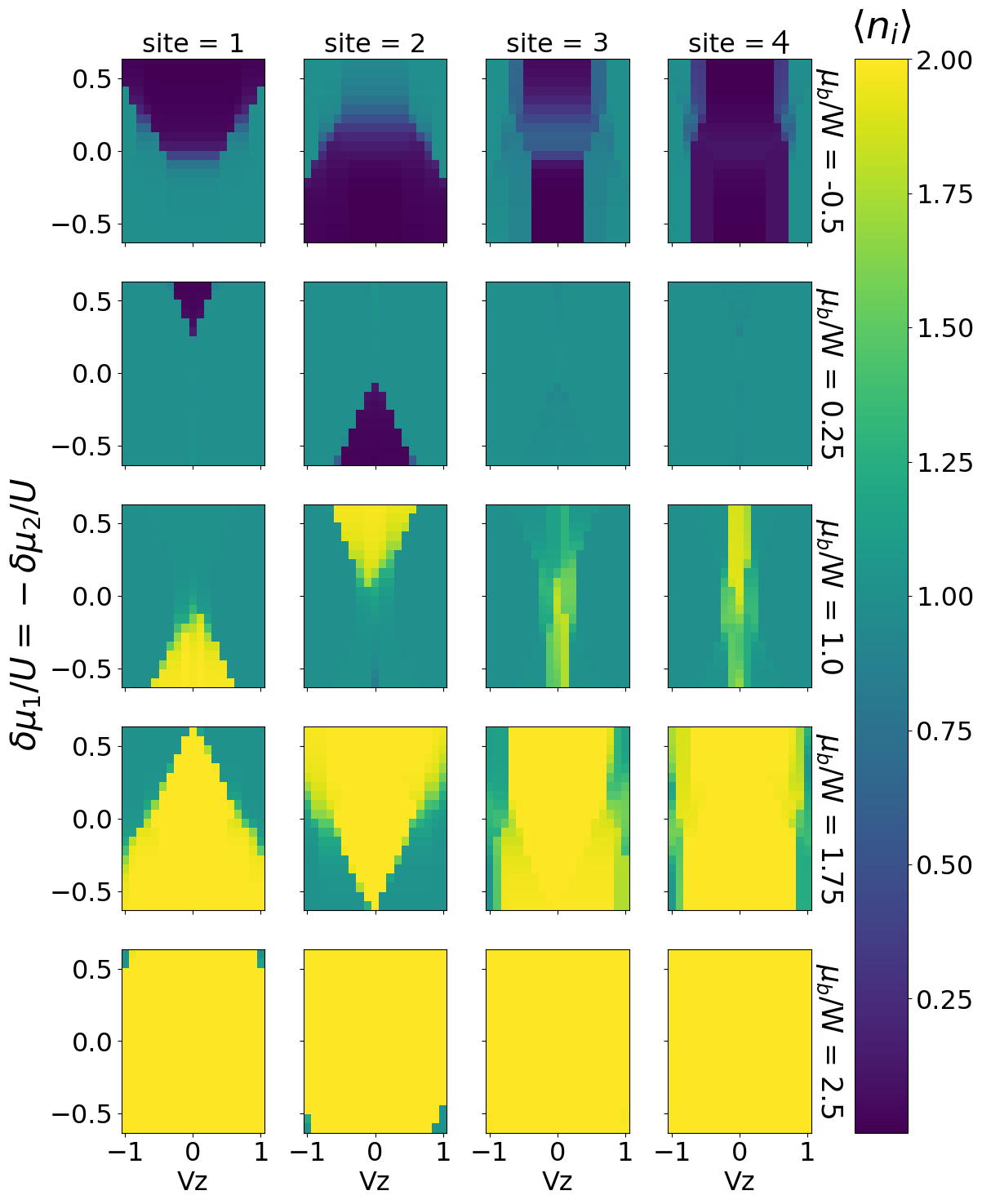}
    \caption{Example of charge stability diagrams used as input for the neural network. The expectation value $\langle n_i \rangle$ is computed while varying $V_z$, $\mu_b$, and $\delta\mu_i=-\delta\mu_{i+1}$ for each site $i$. The resulting charge stability diagram matrix is then fed into the neural network to predict the Hubbard model parameters.}
    \label{fig:ChargeStabilityDiagram}
\end{figure}

The primary goal of this work is to determine the angles $\theta_{ij}$---which determine the strength of the SOC-induced spin-flip hopping processes between nearest-neighbor dots---from charge stability diagrams, which are among the most routinely acquired experimental datasets in quantum dot devices. For the purpose of this work, we first simulate the outcome of such charge stability measurements numerically to produce the training data for the neural network. Specifically, we obtain charge stability diagrams for the quantum dot array described by Eq.~(\ref{eq:hubbard}) via numerical exact diagonalization using the Quspin package~\cite{weinberg2017quspin}. Our simulated measurement outcomes are based on expectation values $\langle n_i\rangle = \text{Tr}(\rho n_i)$, with $\rho = e^{-\beta (H - \sum_i \mu_i n_i)}$, taken across a range of site-dependent chemical potentials $\mu_i= \mu_b + \delta\mu_i$, which can be experimentally controlled by gate voltages, and a range of out-of-plane magnetic fields $V_z$. Here, $\mu_b$ is a constant uniform background potential and $\delta\mu_i$ is a local site-dependent deviation on dot $i$. We work at a low temperature ${k_B}/{\beta} = 0.005 \langle U \rangle$ and retain the lowest $10$ energy eigenstates for the calculation of $\rho$, which is sufficient to account for thermal effects at such low temperatures. All sites are measured simultaneously, yielding the occupation vector $\textbf{n} = (\langle n_1\rangle, \cdots, \langle n_4\rangle)$. These measurements form the charge stability diagrams that will serve as the input for our machine learning algorithm, see Fig.~\ref{fig:ChargeStabilityDiagram} for an example.

\textit{Machine learning setup.\textemdash}%
Each disordered realization of our quantum dot array is represented by a series of randomly generated model parameters $Y$:
 \begin{equation}
Y=\begin{bmatrix} \vec{\theta} & \vec{\epsilon} & \vec{V} & \vec{t} & \vec{U} & \vec{\phi}\end{bmatrix},
\end{equation}
where the vector $\vec{p}$ lists the value of a model parameter $p$ for all sites or all bonds of the lattice (depending on the type of parameter). The array $Y$ is then fed into an exact diagonalization scheme to generate charge stability diagrams as described above. Specifically, for each $Y$, we generate charge stability data for a series of measurement configurations $l=(\mu_b,\delta\mu_i,V_z)$ corresponding to different values of dot chemical potentials and out-of-plane magnetic fields. The total dataset for the device $Y$ is then composed of a series of occupation expectation values on dots 1--4 for the different values of $l$, forming an input matrix X:
\begin{equation}
X=[\textbf{n}^{l=1}, \textbf{n}^{l=2},\dots].
\end{equation}
Specifically, we vary the background chemical potential $\mu_b$ between $[-0.5W, 2.5W]$ with 5 steps. Here, $W = \frac{1}{4}\left(\sum_i \langle U_i \rangle + \langle V_{i,i+1} \rangle\right)$ is a fixed normalization factor that does not vary by sample and can, in principle, be set to unity. Furthermore, for each bond connecting the nearest-neighbor sites $i$ and $j$, we consider a symmetric detuning $\delta\mu_i = -\delta\mu_j$, which is varied between $[-0.6\langle U\rangle, +0.6\langle U\rangle]$ with 15 steps, while we set $\delta\mu_{k \neq i, j} = 0$.  The Zeeman splitting $V_z$ is varied between $[-1, 1]$ with 19 steps.

The matrix $X$ is utilized by the neural network to determine $\vec{\theta}$ as well as the other Hubbard model parameters. Thus, the neural network is trained to perform
\begin{equation}
f_{ML}(X)=Y.
\end{equation}
Unless specified otherwise, we draw the Hamiltonian parameters from the following ranges:
$\theta_{ij} \in [0, 0.2]$, $\epsilon_i \in [-0.5, 0.5]$, $V_{ij} \in [0.0, 0.4]$, $U_i \in [3, 5]$, $t_{ij} \in [0.1, 0.5]$, and $\phi_{ij}\in[0,\pi/2]$. We randomly alternate (with equal probability) between uniform and normal distributions for $U_i$, $V_{ij}$, and $\epsilon_i$ to prevent hard boundaries at the edges of parameter space, while always using a uniform distribution for $t_{ij}$ and $\theta_{ij}$. (For the Gaussian distribution, the parameter ranges specified above represent the $1\sigma$ range with the average located at the center.) We further fix $t_{12} = 0.25$ to set an overall energy scale.

\textit{Neural network.\textemdash}%
The neural network consists of a sequence of 3D vision transformers that take as input expectation values arranged in a tensor with axes $[\mathrm{site}, \delta \mu_i, V_z, \mu_b \otimes \mathrm{pair}]$ and shape $[4, 15, 19, 4 \times 4]$. The site axis, corresponding to the index $j$ in $\langle n_j \rangle$, is treated as the channel dimension, while $V_z$, $\delta \mu_i$, and $\mu_b \otimes \mathrm{pair}$ are treated as the spatial dimensions. Here, $\mu_b \otimes \mathrm{pair}$ denotes the chemical-potential axis stacked with the pair axis, analogous to the index $i$ in $\delta \mu_i$. 

The training process for this neural network must be performed quite slowly in many of the neural network architectures we attempted. In particular, for convolutional-based architectures, we found that for a large number of initial epochs the network becomes trapped at $R^2 \approx 0$, seemingly stuck at predicting the average. This issue is significantly remedied through the use of a 3D vision transformer, see the SM for the specifics of the neural network.
We found it optimal to train the neural network individually for each parameter type, 
which prevents the network from prioritizing parameters with wider bounds. Moreover, there is no reason to assume that the visual features critical to predicting one parameter would be the same for all other parameters.

\begin{figure}[t]
    \centering

    \begin{subfigure}[t]{0.49\linewidth}
        \centering \captionsetup{justification=raggedright,singlelinecheck=off,skip=-4cm,margin={-0.05cm,0cm}}
        \includegraphics[width=\textwidth]{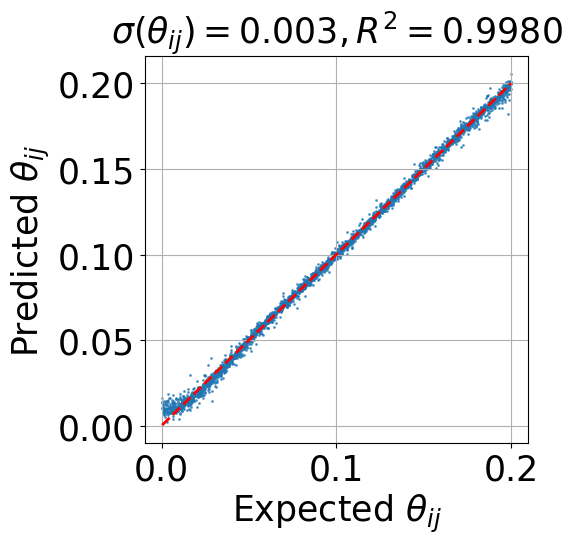}
        \caption{}
    \end{subfigure}
    \hfill
    \begin{subfigure}[t]{0.49\linewidth}
        \centering    \captionsetup{justification=raggedright,singlelinecheck=off,skip=-4cm,margin={-0.05cm,0cm}}
        \includegraphics[width=\textwidth]{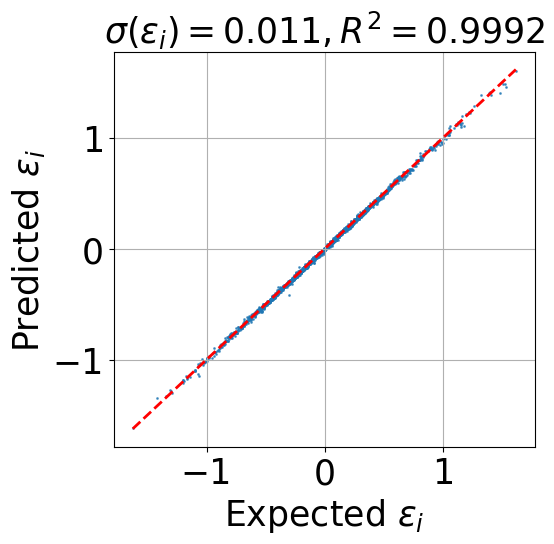}
        \caption{}
    \end{subfigure}
    \caption{Scatter plot showing predicted vs. expected parameters (all bonds/sites) when only $\theta_{ij}$ and $\epsilon_i$ are unknown. (a) $\theta_{ij}$, (b) $\epsilon_{i}$.}
    \label{fig:simpleresult}
\end{figure}

\textit{Results.\textemdash}%
We start by considering a very simple scenario where only $\theta_{ij}$ and $\epsilon_i$ are unknown, while $U_i$, $V_{ij}$, and $t_{ij}$ are known. This allows us to test whether it is fundamentally possible for our neural network to predict a spin-dependent quantity such as $\theta_{ij}$ from the charge stability data we provide as input. For now, we set $\phi_{ij}=0$ for all bonds, while the more general case with random $\phi_{ij}$ will be discussed below. As specified above, we vary $\theta_{ij} \in [0, 0.2]$ and $\epsilon_i \in [-0.5, 0.5]$ (alternating between Gaussian and uniform distributions). We find that our neural network can predict $\theta_{ij}$ nearly perfectly for our four-dot system, with $\sigma(\theta_{ij}) = 0.003$ and $R^2(\theta_{ij}) = 0.9980$, see Fig.~\ref{fig:simpleresult}(a). Here, we use $\sigma(p)$ to denote the standard deviation error (square root of the mean square error) in the prediction of a parameter $p$, and $R^2(p)$ is the standard scaled fidelity that measures the accuracy of the prediction~\cite{Pedregosa2011}.

Furthermore, we also verify that a nonzero $\theta_{ij}$ does not hinder the network's ability to predict $\epsilon_i$. Indeed, we find that $\epsilon_i$ can be predicted very accurately, with $\sigma(\epsilon_i) = 0.011$ and $R^2(\epsilon_i) = 0.9992$, see Fig.~\ref{fig:simpleresult}(b). This shows that neural networks can still be used to aid in the automatic tuning of quantum dot arrays~\cite{durrer2020automated,schuff2024fully,taylor2025neural} even in the presence of spin-flip hopping terms.

\begin{figure}[t]
    \centering

    \begin{subfigure}[t]{0.48\linewidth}
        \centering        \captionsetup{justification=raggedright,singlelinecheck=off,skip=-4cm,margin={-0.05cm,0cm}}
        \includegraphics[width=\textwidth]{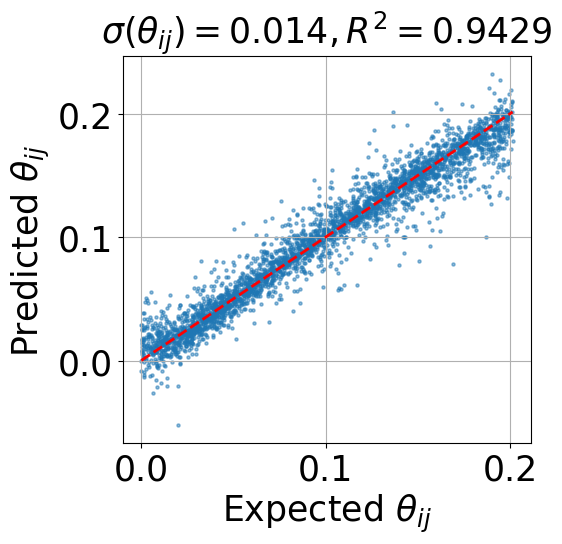}
        \caption{}
    \end{subfigure}
    \hfill
    \begin{subfigure}[t]{0.48\linewidth}
        \centering                \captionsetup{justification=raggedright,singlelinecheck=off,skip=-4cm,margin={-0.05cm,0cm}}
        \includegraphics[width=\textwidth]{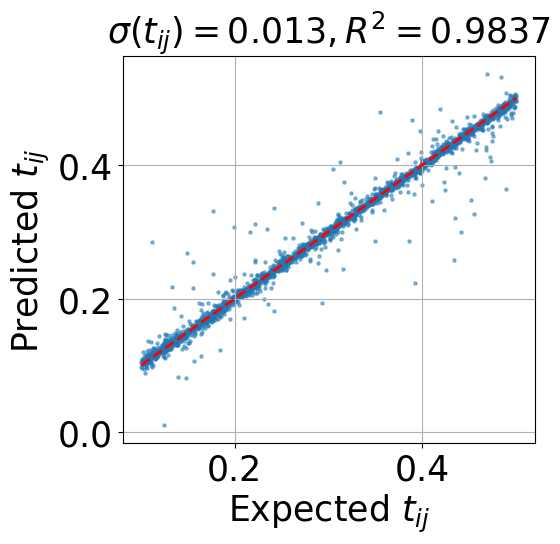}
        \caption{}
    \end{subfigure}

    \caption{Scatter plot showing predicted vs. expected parameters (all bonds) when all model parameters except $\phi_{ij}=0$ are unknown. (a) $\theta_{ij}$, (b) $t_{ij}$.}
    \label{fig:simpleresult2}
\end{figure}

It is natural to ask whether the network can still accurately predict $\theta_{ij}$ even when all Hubbard model parameters are unknown. In particular, the case of unknown hopping amplitudes $t_{ij}$ is interesting from a technical point of view: Since the $t_{ij}$ enter the spin-flip term in Eq.~(\ref{eq:hubbard}) multiplicatively, it is possible that the network learns an effective $t_{ij}$ when predicting $\theta_{ij}$, and that an unknown $t_{ij}$ would therefore significantly degrade our algorithm's performance. However, we find that this is not the case and that the neural network remains capable of making high-fidelity predictions for $\theta_{ij}$ with $R^2(\theta_{ij}) > 0.94$ even when $t_{ij}$ is unknown, see Fig.~\ref{fig:simpleresult2}(a). Although the prediction for the bond connecting dots $1$ and $2$, for which $t_{12} = 0.25$ was pinned, is better than for the other sites [$R^2(\theta_{12})=0.96$], unknown hopping amplitudes $t_{ij}$ only slightly reduce the prediction fidelity. We also note that, while $\theta_{ij}$ was drawn from a limited range $\theta_{ij}\in[0,0.2]$ here to test whether the SOC strength can be predicted in principle, we show in the SM that considering a larger range of $\theta_{ij}\in [0,\pi/2]$ does not significantly affect our results.

Moreover, $t_{ij}$ itself can also be predicted with high fidelity, see Fig.~\ref{fig:simpleresult2}(b). As such, our machine learning algorithm is capable of extracting the complete spin-flip interdot hopping amplitudes $t_{ij}\sin{\theta_{ij}}$ from the provided charge stability data. Additionally, we find very high prediction fidelities $R^2>0.99$ for the remaining Hubbard model parameters $U_i$, $V_{ij}$, and $\epsilon_i$ (not shown here).

Finally, we consider the case where the spin-orbit axes $\vec{q}_{ij}$ are unknown. We incorporate this into our model by allowing $\phi_{ij}$ to vary randomly in $[0, \pi/2]$, while keeping all other parameters varying as before. Indeed, in Ge quantum wells, the dominant SOC is predicted to be of cubic Rashba type~\cite{Winkler2003,Moriya2014,Mizokuchi2017,Marcellina2017,Terrazos2021,Xiong2022,Jirovec2022}, with the spin-orbit axes $\vec{q}_{ij}$ strongly depending on the local electrostatics and the orientation of the quantum dot array with respect to the crystal axes. Furthermore, spin-orbit fields consistent with Dresselhaus-type SOC have also been reported~\cite{Seidler2025}, which can potentially be explained by symmetry breaking at the Ge/SiGe interface~\cite{Rodriguez2023}. As such, the exact spin-orbit axes are generally strongly sample-dependent and non-trivial to predict theoretically.

\begin{figure*}[t]
    \centering
    \begin{subfigure}[t]{0.195\textwidth}
        \centering                        \captionsetup{justification=raggedright,singlelinecheck=off,skip=-3.4cm,margin={-0.05cm,0cm}}
        \includegraphics[width=\textwidth]{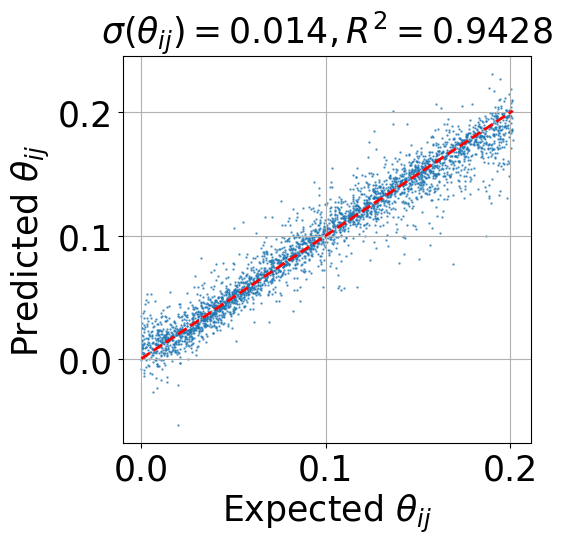}
        \caption{}
    \end{subfigure}
    \begin{subfigure}[t]{0.195\textwidth}
        \centering        \captionsetup{justification=raggedright,singlelinecheck=off,skip=-3.4cm,margin={-0.05cm,0cm}}
        \includegraphics[width=\textwidth]{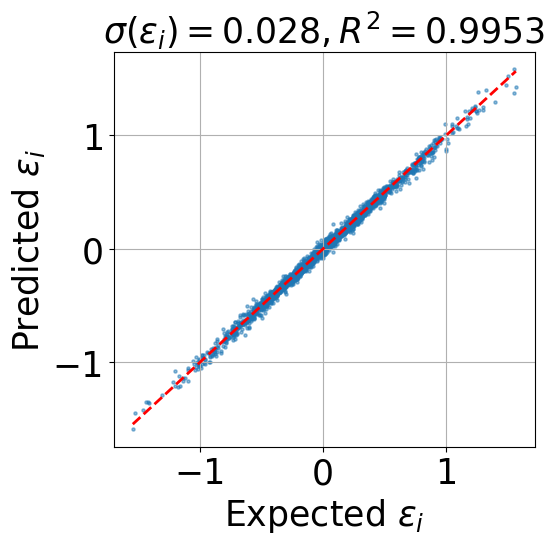}
        \caption{}
    \end{subfigure}
    \begin{subfigure}[t]{0.195\textwidth}
        \centering       \captionsetup{justification=raggedright,singlelinecheck=off,skip=-3.4cm,margin={-0.05cm,0cm}}
        \includegraphics[width=\textwidth]{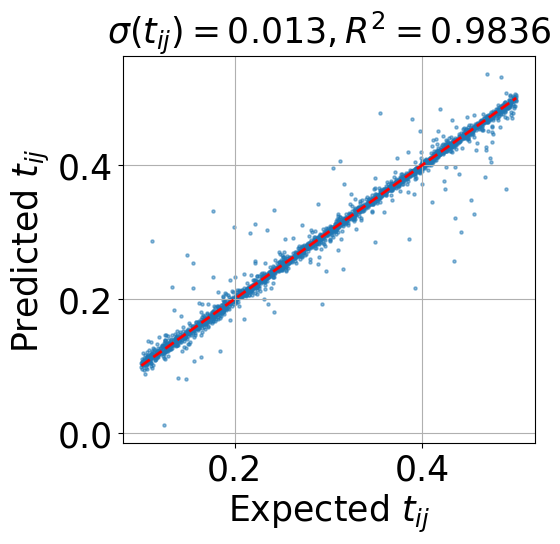}
        \caption{}
    \end{subfigure}
    \begin{subfigure}[t]{0.195\textwidth}
        \centering        \captionsetup{justification=raggedright,singlelinecheck=off,skip=-3.4cm,margin={-0.1cm,0cm}}
        \includegraphics[width=\textwidth]{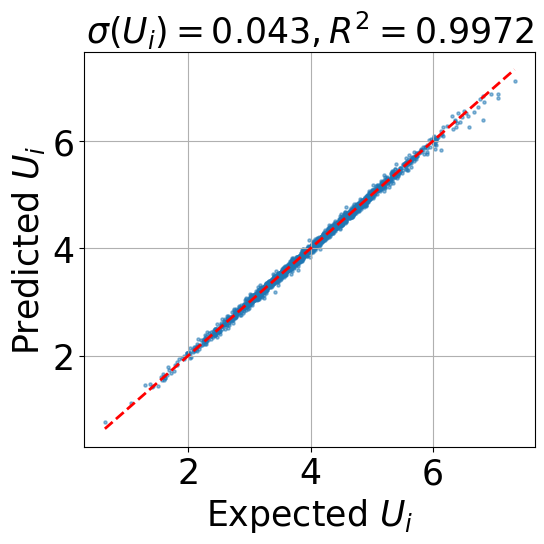}
        \caption{}
\end{subfigure}
    \begin{subfigure}[t]{0.195\textwidth}
        \centering        \captionsetup{justification=raggedright,singlelinecheck=off,skip=-3.4cm,margin={-0.05cm,0cm}}
        \includegraphics[width=\textwidth]{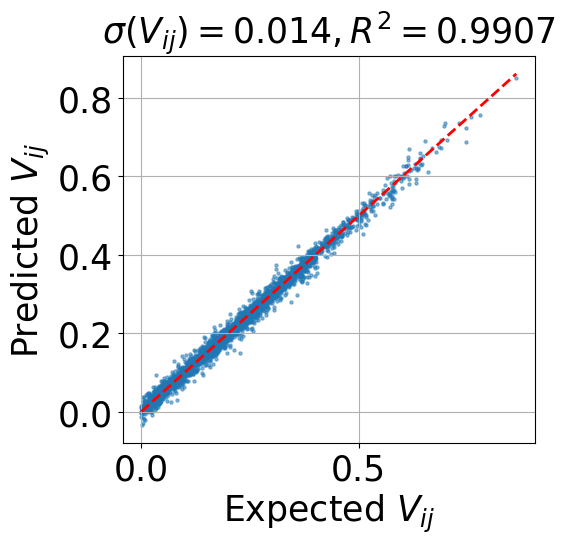}
        \caption{}
\end{subfigure}
    \caption{Scatter plot showing predicted vs. expected model parameters when all parameters, including $\phi_{ij}$, are unknown. (a) $\theta_{ij}$, (b) $\epsilon_{i}$, (c) $t_{ij}$, (d) $U_{i}$, and (e) $V_{ij}$.}    \label{fig:allresult}
\end{figure*}

Fortunately, we find that the unknown $\phi_{ij}$ does not affect our ability to predict $\theta_{ij}$, as evidenced by a high fidelity of the test-set $\sigma(\theta_{ij}) = 0.014$ and $R^2(\theta_{ij}) = 0.94$, see Fig.~\ref{fig:allresult}(a). This represents effectively no reduction compared to the $\phi_{ij} = 0$ case (which essentially corresponds to the case of pure linear Rashba SOC). The fact that $\phi_{ij}$ is unknown also does not significantly hinder the network's ability to predict the other Hubbard model parameters; see Figs.~\ref{fig:allresult}(b-e) and Tab.~\ref{Tab:PAllResults}.

\begin{table}[b]
\centering
\label{tab:results}
\begin{tabular}{lccc}
\hline
\textbf{Parameter} & Range & $\sqrt{\textbf{MSE}}$ & \textbf{$R^2$} \\
$\theta_{ij}$ & [0,0.2] & 0.014 & 0.9428\\
$t_{ij}$ & [0.1,0.5] & 0.013 & 0.9836\\
$\epsilon_{i}$ & [-0.5,0.5] & 0.028 & 0.9953 \\
$U_{i}$ & [3,5] & 0.043 & 0.9972 \\
$V_{ij}$ & [0.0,0.4] & 0.014 & 0.9907 \\
$\phi_{ij}$ & $[0,\pi/2]$ & Failed & 0.0\\
\hline
\end{tabular}
\caption{Model performance and data ranges when all parameters, including $\phi_{ij}$, are unknown. Fidelity is shown in terms of standard deviation error (square root of mean square error) and $R^2$. The prediction of $\phi_{ij}$ fails as the corresponding effect on the charge stability diagrams is too small to be discerned by the neural network. The indicated parameter ranges are exact when a uniform distribution is used, whereas they represent the $1\sigma$ range (average located at the center) when we use a Gaussian distribution.}
\label{Tab:PAllResults}
\end{table}

On the other hand, our network is unable to predict $\phi_{ij}$ itself from the provided charge stability diagrams. Despite achieving $R^2(\phi_{ij}) > 0.9999$ on the training set, the test set remains at a best value of $R^2(\phi_{ij}) \approx 0$, which is equivalent to returning the average. This is a strong indicator that the information required to predict $\phi_{ij}$ is either not contained in the charge stability diagrams at all or completely washed out by other effects. The high training fidelity then arises from the network memorizing the data rather than learning a generalizable feature.

In our case, $\phi_{ij}$ is not predictable because its quantitative effect on the charge stability diagrams is on the order of $\delta \langle n\rangle=10^{-3}$, which is much less than what the neural network can distinguish in the presence of disorder. This is further supported by the fact that we can predict $\phi_{ij}$ with high fidelity when all other disorder sources are removed. In the SM, we show that $\phi_{ij}$ becomes predictable even in the presence of disorder when the measurement configuration is extended to include a second magnetic field direction in the $xz$-plane.

\textit{Conclusion.\textemdash}%
We have described a machine learning method to determine the spin-flip interdot hopping amplitudes---and, thus, the strength of the effective spin-orbit coupling---in semiconductor hole quantum dot arrays from charge stability data. We find that standard charge stability diagrams measured in a varying out-of-plane magnetic field contain sufficient information to accurately predict the spin-flip hopping amplitudes with high fidelity.

In addition to the spin-orbit coupling strength, our neural network can also predict the remaining Hubbard model parameters with high fidelity, thus offering a simple and automatic method to obtain a quantitative theoretical description (in terms of a spin-orbit coupled Hubbard model) of a given quantum dot array from experimentally accessible data. As such, our machine-learning method can help improve the current experimental characterization and tuning process of hole quantum dot devices. We emphasize that our algorithm processes data from all $4$ dots in the $2\times 2$ array simultaneously, such that crosstalk effects are automatically included.

Finally, we note that while we have focused on a $2\times 2$ Ge hole quantum dot array as a concrete and experimentally relevant example~\cite{Lawrie2020,Hendrickx2021,Lawrie2023}, our method is equally applicable to other array geometries as well as to hole spin qubit systems in other materials such as, e.g., Si~\cite{Fang2023}.

\textit{Acknowledgment.\textemdash}%
This work is supported by the Laboratory for Physical Sciences. We also thank UMD HPC Zaratan for computational resources provided. 
\bibliography{mainbib}
\appendix
\clearpage
\renewcommand{\thesection}{\Roman{section}}
\vspace{3cm}
\onecolumngrid
\begin{center}
    {\bf \large Supplementary Material for ``Predicting spin-orbit coupling in Ge hole spin qubit arrays with vision-transformer-based neural networks''}
\vspace{1cm}
\end{center}
\twocolumngrid
\setcounter{page}{1}
\setcounter{secnumdepth}{3}
\setcounter{equation}{0}
\setcounter{figure}{0}
\renewcommand{\theequation}{S-\thesection.\arabic{equation}}

\renewcommand{\thefigure}{S\arabic{figure}}
\renewcommand\figurename{Supplementary Figure}
\renewcommand\tablename{Supplementary Table}

\section{Extended neural network details}
\begin{figure}[tb]
    \centering
    \hfill
    \begin{subfigure}[b]{0.49\linewidth}
        \centering
        \includegraphics[width=\textwidth]{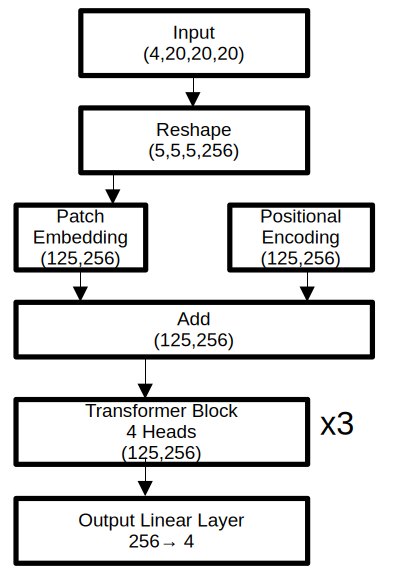}
        \caption{}
    \end{subfigure}
    \begin{subfigure}[b]{0.49\linewidth}
        \centering
        \includegraphics[width=\textwidth]{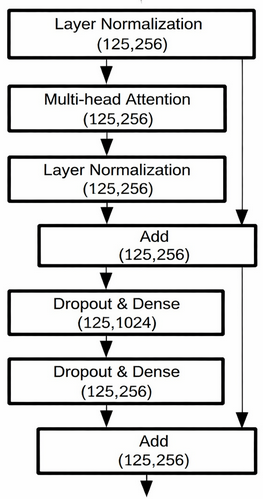}
        \caption{}
    \end{subfigure}
    \caption{(a) Diagram of the full neural network, which uses a standard vision transformer setup as described in the text. The input is split into patches and then converted to embedding space with positional information included additively through sinusoidal functions. Three blocks of transformers with four heads each are applied to the input, and then fed into a linear layer for final output. The final output is a vector of values for a single disorder source at each site/bond. (b) Transformer block diagram.}    \label{fig:NN}
\end{figure}

The neural network consists of three layers of 3D transformers, where the input is embedded into patches of $4\times 4$ images with four frames each. We make use of the simplified improved vision transformer from Ref.~\cite{beyer2022better} using sinusoidal positional embedding. The axes represent [site, $\delta \mu_i$, $V_z$, $\mu_b \otimes \text{pair}$]. The network input additionally has four channels corresponding to the different sites of the system. The transformers make use of an embedding (mapped through a linear layer) of 512 with 4 heads, and the perceptron layer has a total of 1024 neurons each. These parameters were chosen through trials to be optimal for $\theta_{ij}$ prediction; however, separate neural networks with the same architecture are used for the prediction of the additional Hubbard model parameters. The neural network was trained for 500 epochs or until convergence, whichever was first. The epochs were run with a gradually decreasing learning rate (starting at 1E-3 to 1E-5) and weight decay (1E-4 to 1E-6). A figure showing the neural network can be seen in Fig. \ref{fig:NN}. We made use of the Pytorch \cite{paszke2019pytorch} and VIT-Pytorch packages~\cite{wang_vit_pytorch}.

\section{$g$-tensor variations}

For gate-defined Ge hole quantum dots in Ge/SiGe heterostructures, the effective $g$-tensor typically varies from dot to dot due to small differences in strain and the local electrostatic landscape. To assess whether an unknown site-dependent $g$-tensor $\overleftrightarrow{g_i}$ affects our neural network's ability to predict $\theta_{ij}$, we replace the magnetic term $\propto V_z$ in Eq.~(1) by a more general expression
\begin{align}
H_{B}&=\sum_{i,\alpha,\alpha'} \,c^\dagger_{i\alpha}(\vec{B}\cdot\overleftrightarrow{g_i}\cdot \vec{\sigma})^{\alpha\alpha'}c_{i\alpha'}.
\end{align}
As in the main text, we focus on an out-of-plane magnetic field $\vec{B}=(0,0,B_z)$, such that only the three components $g^{xz}_i,g^{yz}_i,g^{zz}_i$ of the $g$-tensor enter our Hamiltonian. Here, the off-diagonal components $g^{xz}_i,g^{yz}_i\ll g^{zz}_i$ are small but generally nonzero, such that the magnetic $z$ axes of the dots are generally slightly tilted with respect to the device $z$ axis. In the following, we parametrize the vector $\vec{g}_i:=(g^{xz}_i,g^{yz}_i,g^{zz}_i)$ by its length $|\vec{g}_i|$ as well as two angles $\theta_g^i$ and $\phi_g^i$, giving us $\vec{g}_i=|g_i|(\sin\theta_g^i\cos\phi_g^i,\sin\theta_g^i\sin\phi_g^i,\cos\theta_g^i)$. We then include dot-dependent $g$-tensor variations by randomly selecting $|g_i|\in [0.9,1.1]$, $\phi^i_g\in[0,\pi/2]$, and $\theta^i_g\in[0,\delta \theta_g]$ for each dot. We allow for a large range of variations in the azimuthal angle $\phi_g^i$, while we consider three different cases $\delta \theta_g=0.05, 0.1, 0.2$ for the polar angle $\theta_g^i$, which is often found to be very small ($\lesssim$ a few degrees)~\cite{Hendrickx2024,Seidler2025}. 
\begin{table}[tb]
\centering
\label{tab:results}
\begin{tabular}{lccc}
\hline
$\delta \theta_g$ & $\sqrt{\mathrm{MSE}(\theta_{ij})}$ & \textbf{$R^2(\theta_{ij})$} \\
 0.05 & 0.016 &0.9223 \\
 0.1 & 0.021 &0.8620 \\
 0.2 & 0.033 &0.6703\\
\hline
\end{tabular}
\caption{Model performance when $g$-tensor variations are included (all other parameters as in the main text) for different ranges of polar tilt angles $[0,\delta\theta_g]$. Fidelity is shown in terms of standard deviation error (square root of mean square error) and $R^2$. }
\label{Tab:}
\end{table}
\begin{figure*}[tb]
    \centering
    \begin{subfigure}[t]{0.25\linewidth}
        \centering        \includegraphics[width=\textwidth]{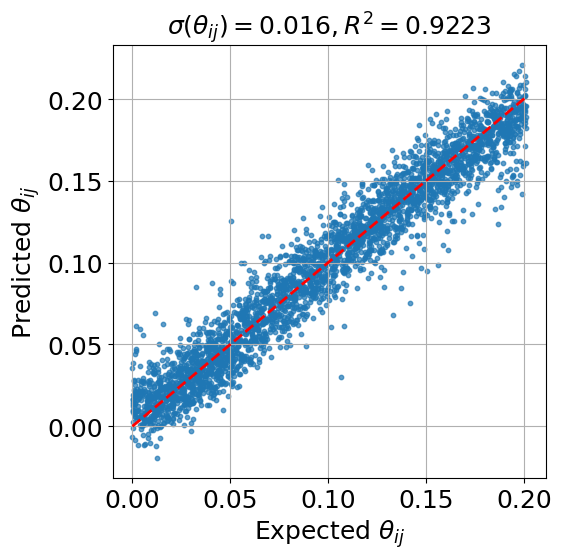}
        \caption{}
    \end{subfigure}
    \hspace{0.8cm}
    \begin{subfigure}[t]{0.25\linewidth}
        \centering
        \includegraphics[width=\textwidth]{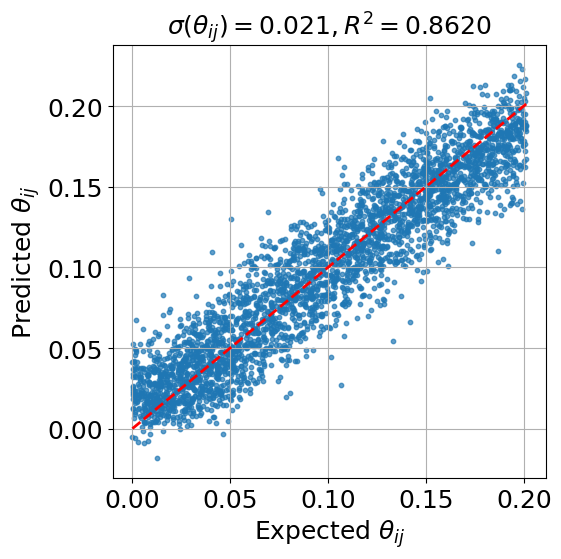}
        \caption{}
    \end{subfigure}
    \hspace{0.8cm}
    \begin{subfigure}[t]{0.25\linewidth}
        \centering
        \includegraphics[width=\textwidth]{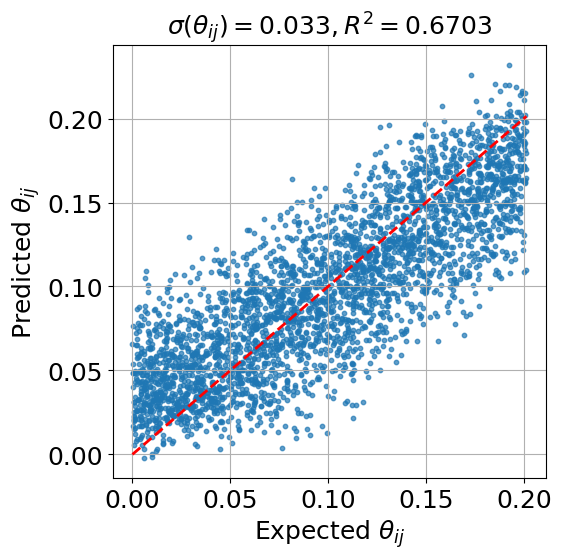}
        \caption{}
    \end{subfigure}
    \caption{Scatter plot of predicted vs. expected $\theta_{ij}$ results when $\vec{g}_{i}$ disorder is included. Each subplot refers to a different magnitude $\theta_g$ disorder. (a) $\delta \theta_g=0.05$, (b) $\delta \theta_g=0.1$, (c) $\delta \theta_g=0.2$.}    \label{fig:sallresult}
\end{figure*}
Table~\ref{Tab:} shows that $R^2(\theta_{ij})$ decreases as the variability in $\theta_g^i$ increases, but nevertheless remains  high for experimentally realistic tilt angles ($\sim$ first row of Table~\ref{Tab:}), see also Fig.~\ref{fig:sallresult}. Additionally, we find that the variation in $|g_i|$ has a negligible effect on $R^2(\theta_{ij})$. Finally, we have also checked that $g$-tensor variations have no effect on the predictability of the other Hubbard model parameters, as is expected since these quantities are spin-independent.

We note that it would be interesting to train the neural network on even more diverse datasets including, e.g., gate-voltage-dependent $g$-tensors to bring the training data closer to the experimental reality. This is left to future work. Generally, while our work presents a minimal working example for the machine-learning-based characterization of hole spin qubit arrays, various extensions could straightforwardly be implemented if greater computational resources are used.

\section{Additional results for a larger range of $\theta_{ij}$}

\begin{figure}[b]
    \centering
    \begin{subfigure}[t]{0.48\linewidth}
        \centering        \includegraphics[width=\textwidth]{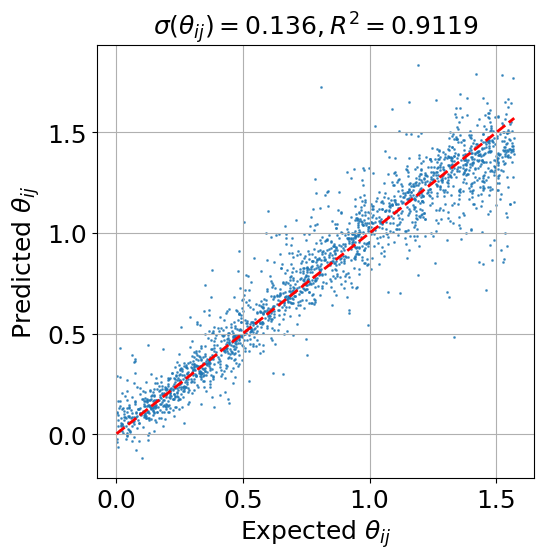}
        \caption{}
    \end{subfigure}
    \hfill
    \begin{subfigure}[t]{0.48\linewidth}
        \centering
        \includegraphics[width=\textwidth]{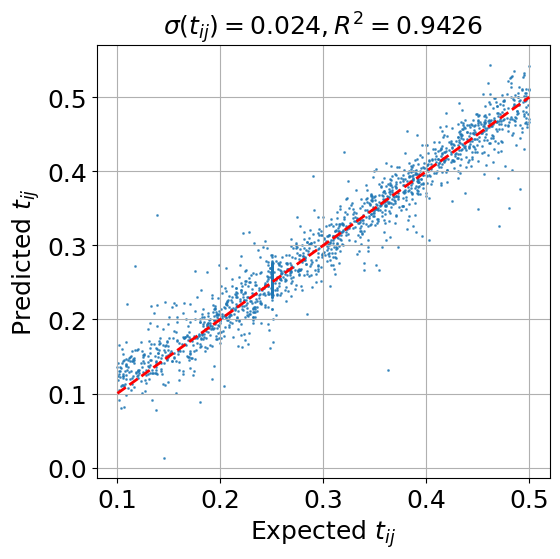}
        \caption{}
    \end{subfigure}
    \caption{Scatter plot showing predicted vs. expected parameters for (a) $\theta_{ij}$ and (b) $t_{ij}$ (all bonds) when the range of spin-flip angles is extended to $\theta_{ij}\in[0,\pi/2]$. The prediction fidelities for $\theta_{ij}$ and $t_{ij}$ are only marginally lower compared to the main text, while all other parameters (not shown here) are unaffected.}    \label{fig:sallresult2}
\end{figure}

In the main text, we chose the range $\theta_{ij}\in[0,0.2]$ to demonstrate that our neural network can predict the spin-orbit coupling strength from charge stability diagrams. However, experimentally, larger values of $\theta_{ij}$ are possible. Therefore, in Fig.~\ref{fig:sallresult2}, we present additional results showing that $R^2(\theta_{ij})$ is only marginally reduced if $\theta_{ij}$ is varied over a much larger range $\theta_{ij}\in[0,\pi/2]$. Similarly, the prediction fidelities for the other Hubbard model parameters remain high as well. As such, our machine-learning method does not depend sensitively on the range $\theta_{ij}$ is chosen from.

\section{Prediction of $\phi_{ij}$ using multiple magnetic field axes}
\begin{figure}[tb]
        \centering
        \includegraphics[width=0.5\linewidth]{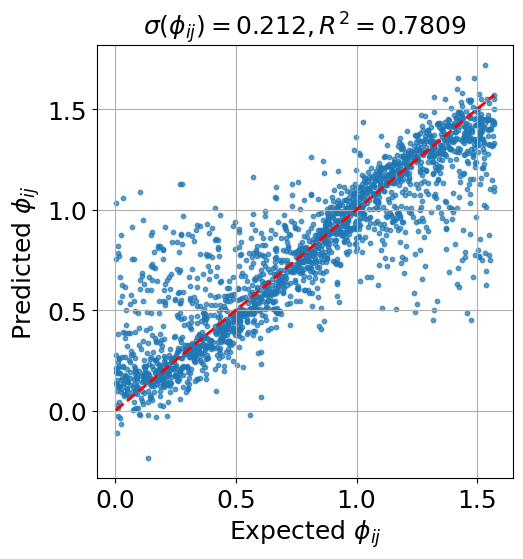}
    \caption{Scatter plot showing predicted vs. expected $\phi_{ij}$ (all bonds) when the range of spin-flip angles is extended to $\theta_{ij}\in[0,\pi/2]$ and an additional full $\vec{B}=(B/\sqrt{2},0,B/\sqrt{2})$ axis is included.}    \label{fig:2axisresult}
\end{figure}

We also evaluated whether $\phi_{ij}$ could be predicted over the broader range $\theta_{ij}\in[0,\pi/2]$ by incorporating information from a second magnetic-field axis during neural-network training. In this case, we considered fields applied along the $z$ axis together with a secondary axis oriented at an angle $\pi/4$ in the $xz$ plane $\vec{B}=(B/\sqrt{2},0,B/\sqrt{2})$, using the full set of 19 magnetic-field values for both directions. Under this configuration, the network was able to predict $\phi_{ij}$ with moderately high fidelity, achieving $R^2 = 0.78$ (see Fig. \ref{fig:2axisresult}). This predictive power arises because, in the larger $\theta_{ij}$ case, the effect of $\phi_{ij}$ is less washed-out, while the second field axis allows the neural network to filter the information from the first. Together, these effects allow the contribution of $\phi_{ij}$ to be more clearly isolated and extracted.

\end{document}